\documentclass[aps,showpacs,twocolumn,showkeys,amsmath,amssymb,floatfix,nofootinbib,preprintnumbers]
{revtex4}
\usepackage{bm}
\usepackage{epsfig}

\begin{document}

\title{Diquark Substructure in $\phi$ Photoproduction}

\author{Richard F. Lebed}
\email{richard.lebed@asu.edu}
\affiliation{Department of Physics, Arizona State University, Tempe,
Arizona 85287-1504, USA}

\date{October, 2015}

\begin{abstract}
  Observed enhancements in the forward and backward directions for
  $\phi$ meson photoproduction off nucleons are shown to be
  explainable by the production of a nonresonant recoiling $(su)$
  diquark, $(\bar s ud)$ triquark pair.  We show that the necessity of
  maintaining approximate collinearity of the quarks within these
  units constrains configurations with the minimum momentum transfer,
  and hence maximal amplitudes, to lie preferentially along the
  reaction axis.
\end{abstract}


\maketitle


\section{Introduction} \label{sec:Intro}

LHCb has recently observed~\cite{Aaij:2015tga} two exotic states,
$P_c^+(4380)$ and $P_c^+(4450)$, at high significance in the $J/\psi
\, p$ spectrum of $\Lambda_b \to J/\psi \, K^- p$.  In addition to
extracting the state masses and widths, the collaboration also
measured the phases of the production amplitude and found them to be
compatible with genuine resonant behavior.  Such properties strongly
support the assertion that true pentaquark states have at last been
revealed.  Should the $P_c^+$ states be confirmed at another
experiment, they will join $X$, $Y$, and $Z$ exotic mesons (widely
believed to have valence tetraquark structure) as the first hadrons
observed at high significance to lie outside of the textbook $q\bar
q$-meson, $qqq$-baryon paradigm.

Of particular note is that all of the observed exotic states contain
$c\bar c$ (or $b\bar b$) pairs.  The possibility of light-flavor ($u$,
$d$, $s$) tetraquarks or pentaquarks has been discussed for decades,
but despite intense experimental efforts, no unambiguous signal of
such a state has ever survived scrutiny.  Nor has a singly-heavy
hadron ($D_s$, $\Lambda_b$, {\it etc.}) been identified as
unambiguously exotic.  One possible explanation for this first
appearance in doubly-heavy channels, as argued implicitly in
Ref.~\cite{Brodsky:2014xia} and explicitly in
Refs.~\cite{Blitz:2015nra,Lebed:2015tna}, is that the exotic consists
of two components, each of which contains a heavy quark and therefore
is fairly compact, and which are separated from each other in the
sense of having a suppressed large wave function overlap.  Such a
configuration is much more difficult to realize with light quarks.
While this scenario could be achieved by hadronic molecules (for
instance, if the famous $X(3872)$ tetraquark candidate has a $c\bar c
u \bar u$ valence-quark structure organized into a $D^0 \bar D^{*0}$
molecule), in this paper we use the innovation of
Ref.~\cite{Brodsky:2014xia} to assert the existence of hadrons formed
of separating {\em colored\/} components that are individually held
together by the attractive ${\bf 3} \otimes {\bf 3}
\supset \bar {\bf 3}$ color interaction, and collectively prevented
from separating asymptotically far due to color confinement.

If the existence of observable exotics is contingent upon each of its
components containing a heavy quark, then one may hope that some
vestige of the exotic behavior persists in analogous production
mechanisms with valence $s\bar s$ rather than $c\bar c$ ($b\bar b$)
quark pairs.  In Ref.~\cite{Brodsky:2014xia}, the observed tetraquark
states were argued to arise from a rapidly separating
diquark-antidiquark pair, $(cq)_{\bar {\bf 3}} (\bar c \bar
q^\prime)_{\bf 3}$ ($q$ in the following generically indicates $u$ or
$d$ quarks) that subsequently hadronizes only through the large-$r$
tails of meson wave functions stretching from the quarks in the
diquarks to the antiquarks in the antidiquarks.  This picture was
extended in Ref.~\cite{Lebed:2015tna} to explain the pentaquark states
as an analogous diquark-anti{\em triquark\/} pair (this combination
having been first proposed for lighter-quark pentaquarks in
Ref.~\cite{Karliner:2003dt}), $(c u)_{\bf 3} [\bar c (ud)_{\bar{\bf
    3}}]_{\bf 3}$, where the $(ud)$ diquark is inherited from the
parent $\Lambda_b$ baryon, and the triquark is seen to assemble from a
further $\bar {\bf 3} \otimes \bar {\bf 3} \supset {\bf 3}$
attraction.  A short review of these papers appears in
Ref.~\cite{Lebed:2015sxa}.  If this mechanism produces such prominent
effects in the $c\bar c$ system as pentaquark resonances, then one may
hope to see at least a remnant of the mechanism in the $s\bar s$
system, in the form of peculiar features appearing in the data.

Shortly after the LHCb announcement, multiple theoretical papers
appeared, discussing various interpretations of the $P_c^+$ states and
advocating for experiments in which to study them.  Among the latter,
three separate
collaborations~\cite{Wang:2015jsa,Kubarovsky:2015aaa,Karliner:2015voa}
proposed using $\gamma N \to P_c \to J/\psi N^{(*)}$ photoproduction
of nucleons $N$ as sensitive tests of the internal structure of the
$P_c$ states.  In each case, the $c\bar c$ pair arises through the
dissociation of the incoming photon.  By the reasoning of the previous
paragraph, one may therefore ask if any unusual features have arisen
in the analogous $s\bar s$ process of $\phi$ photoproduction, $\gamma
p \to \phi p$.

In fact, a detailed experimental study of $\phi$ photoproduction was
also published fairly recently by the CLAS Collaboration at Jefferson
Lab~\cite{Dey:2014tfa}.  Data in the neutral ($K_L K_S$) mode are
presented in Ref.~\cite{Adhikari:2013ija} The most interesting feature
in the cross section data is a forward-angle ``bump'' structure at
$\sqrt{s} \approx 2.2$~GeV (about 250~MeV above threshold) that rises
about a factor 2--3 above a more mildly varying curve in $\sqrt{s}$.
However, this structure appears only at the most forward angles (in
which $\phi$ lies in the same direction as $\gamma$ in the
center-of-momentum [c.m.]  frame), and as argued in
Ref.~\cite{Dey:2014npa}, such a unidirectional structure almost
certainly does not indicate a resonance.  In addition, the cross
section data indicate a small but clear increase at the most backward
angles (a factor of perhaps 50 smaller than the forward
peak\footnote{M. Dugger, private communication.}); while
Ref.~\cite{Dey:2014tfa} mentions a possible origin for backward
enhancement in the $u$-channel exchange of a nucleon, they also
comment that such a direct $\phi$-$p$ coupling would require a nucleon
strangeness content or a violation of OZI suppression surprisingly
larger than expected.

\begin{figure}[!ht]
\begin{center}
\includegraphics[width=\linewidth]{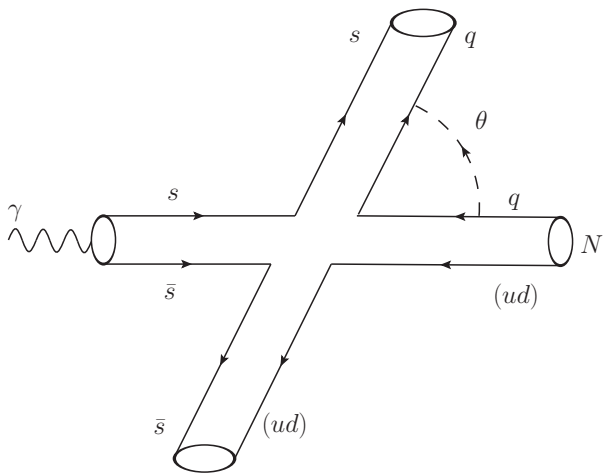}
  \caption{The constituent exchange for $\phi$ photoproduction via
  creation of an $(sq)_{\bar{\bf 3}}$ diquark-$\left[ \bar s
  (ud)_{\bar{\bf 3}} \right]_{\bf 3}$ antitriquark pair.}
\label{Fig:Exchange}
\end{center}
\end{figure}

Here we argue that both features can be explained in a very simple
way: A fraction of $\phi$ photoproduction events proceed through a
constituent exchange between the $\gamma$ (dissociated into an $s
\bar s$ state) and the nucleon [treated as a bound state of a $(ud)$
diquark and a light quark $q$] (Fig.~\ref{Fig:Exchange}).  The
scattering products are a diquark-antitriquark pair, $(s q)_{\bar{\bf
3}} [\bar s (ud)_{\bar{\bf 3}}]_{\bf 3}$, completely analogous to the
construction proposed for the $P_c^+$ resonances, but as experiment
indicates, nonresonant in this case.
That is, the anomalies correspond to ``would-be'' pentaquarks.  The
enhancement in the forward and backward directions, with the forward
direction favored, is explained by the preference of the diquark and
triquark to minimize momentum transfer by aligning with the $\gamma p$
process axis.

This paper is organized as follows: In Sec.~\ref{sec:MomTrans} we
discuss the forward-backward enhancement of hadronic cross sections
driven by constituent-exchange momentum transfers.  A simple model
based upon minimal gluon exchanges is presented in
Sec.~\ref{sec:Model}, and the calculation and results are presented in
Sec.~\ref{sec:Results}.  Section~\ref{sec:Concl} summarizes and
concludes.

\section{Momentum Transfer Enhancement}
\label{sec:MomTrans}

One of the quintessential triumphs of quantum field theory is its
natural prediction of a Rutherford-type differential scattering cross
section, in which the invariant amplitude ${\cal M}$ for the process
$p_1 p_2 \to p_1^\prime p_2^\prime$ (using momenta to label the
particles) scales as the inverse of the 4-momentum transfer $t = q^2 =
(p_1^\prime - p_1)^2 = (p_2 - p_2^\prime)^2$.  A dependence ${\cal M}
\propto 1/(q^2 - m^2)$ arises from the virtual exchange of a quantum
of mass $m$ of the mediating interaction.  For example, in $pp$
elastic scattering the quantum (at least at low energies) may be
considered a single neutral virtual meson, such as $\pi^0$.  At higher
energies, the relevant degrees of freedom exchanged become quarks and
gluons.

Elastic scattering via mediators of negligible mass also features the
characteristic Rutherford angular dependence $1/q^2 \propto 1/ \! \sin^2
\! (\theta/2)$, where $\theta$ is the c.m.\ scattering angle.  Inelastic
$2 \to 2$ scattering (and scattering with heavier mediators) also
contains this factor, although the relation is then no longer a strict
proportionality, and the effect is muted.  In either case, however,
the $\sin^2 \!  (\theta/2)$ factor indicates strong peaking of ${\cal
M}$ in the forward direction, and for precisely the reason that
Rutherford understood: Greater particle deflections demand stronger
forces, requiring scattering events that, for given fixed
initial-particle energies, are comparatively rarer.  If the scattered
particles $p_1^\prime, p_2^\prime$ cannot be unambiguously associated
with the initial particles $p_1, p_2$, respectively---as occurs either
if $p_1, p_2$ represent identical particles, or if the reaction is
sufficiently inelastic that the association is less than
perfect---then one also expects an enhancement in ${\cal M}$
corresponding to the minimization of the 4-momentum transfer $u =
(p_2^\prime - p_1)^2 = (p_2 - p_1^\prime)^2$.  $u$ contains the factor
$\cos^2 (\theta/2)$, which means that ${\cal M}$ also peaks in the
backward direction.

A prominent and historically important example of this
forward-backward peaking occurs in $pn$ elastic scattering.  Indeed,
the strong backward peak, now understood to be due to the
charge-exchange reaction $p \leftrightarrow n$ in which charged mesons
are traded between the nucleons, is one of the best pieces of evidence
for strong interactions respecting fundamental isospin symmetry,
particularly the equivalence of $p$ and $n$ under strong interactions:
In $pn$ scattering the backward peak is almost as high as the forward
peak, because $p$ turns into $n$ and vice versa but remains otherwise
undeflected.  It is interesting to note (as discussed, {\it e.g.}, in
Ref.~\cite{Gibbs:1994um}) that simple one-pion exchange, which one
might naively expect to be responsible for the full peak, is actually
well known to lead to a zero differential cross section at
180$^\circ$.  At the constituent level, a quark exchange $u
\leftrightarrow d$ occurs.

For inelastic processes, the rule of thumb for estimating the relative
height of the forward and backward peaks appears to follow from the
similarity of the initial and final particles.  Taking $p_1$
($p_1^\prime$) to be the lighter initial (final) particle, one expects
the forward peak to be generically higher.  As an explicit example,
consider $K \Lambda$ photoproduction; here, $p_1 = \gamma \to s \bar
s$, $p_1^\prime = K^+$, $p_2 = p$, and $p_2^\prime = \Lambda$.  The
recent CLAS data for this process~\cite{McCracken:2009ra} shows a
backward peak in the differential cross section that is a factor of a
few smaller than the forward peak (the ratio depending strongly upon
$\sqrt{s}$, reaching a maximum of about 1/2 around 2.1~GeV),
supporting the general dynamical picture described here.  Indeed, a
small modification of Fig.~\ref{Fig:Exchange} provides a simple
quark-exchange picture for $K\Lambda$ photoproduction: Just exchange
$s \leftrightarrow \bar s$.  The preference of forward ($\theta \to
0$) scattering indicates the relative dynamical preference for not
diverting the (heavier) baryon line.  Alternately, at the constituent
level for $K \Lambda$ one sees that the lightest constituent $q$ (and
$s$) suffers a backward scattering for $\theta \to 0$, while the
heavier $(ud)$ diquark (and $\bar s$) remains undeflected.
Nevertheless, one expects a smaller enhancement for $\theta \to \pi$
as well, where the momentum transfers of $q$ and $s$ are minimized.

\section{A Simple Diquark-Triquark Model}
\label{sec:Model}

\begin{figure}[!ht]
\begin{center}
\includegraphics[width=\linewidth]{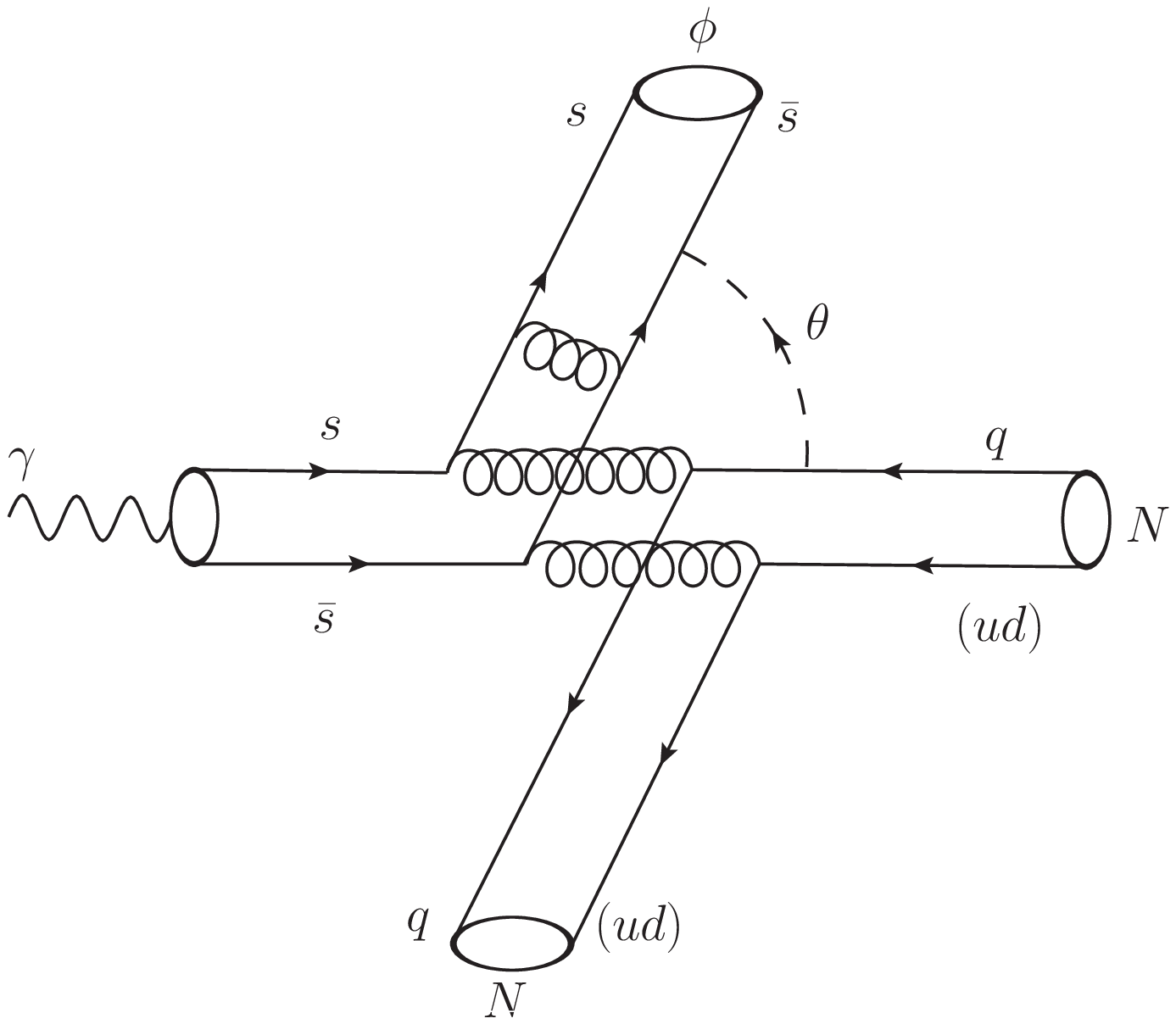}
  \caption{One of many gluon-exchange Feynman diagrams contributing to
  the Pomeron exchange mechanism for $\gamma N \to \phi N$.}
\label{Fig:Pomeron}
\end{center}
\end{figure}

Once the scattered particles in a $2 \to 2$ process are resolved into
constituents, the overall momentum transfer $q^2$ is no longer the
only independent kinematical quantity contributing to angular
dependence.  The dominant diagram for $\gamma N \to \phi N$ is
generally attributed to Pomeron
exchange~\cite{Freund:1967,Barger:1970wk}, in which the $\gamma$
dissociates into the $s\bar s$ pair.\footnote{Alternate diagrams are
possible in which $\gamma$ couples directly to the $N$, and the $s\bar
s$ pair is created from gluons emitted from the struck $N$.  At high
energies, the latter diagram actually appears to be
dominant~\cite{Goritschnig:2014goa}.}  At the level of fundamental
QCD, the corresponding Feynman diagrams include as their most simple
representative Fig.~\ref{Fig:Pomeron}, although the full Pomeron would
include many more gluons, particularly those cross-linking the
exchanged gluon pair in this figure.
The gluon exchanged between the final-state $s\bar s$ pair merely
indicates that, in order for the four constituents to interact, share
momentum, and be diverted to their final directions, a minimum of
three gluons must be exchanged\footnote{Note that the $(ud)$ is
  treated as a single unit in the scattering.}.  In fact, the complete
diagrams are even more complicated than indicated because we have
represented the nucleon $N$ as a bound pair of a light quark $q$ and a
diquark $(ud)$; indeed, the gluons can couple separately to either of
the quarks in $(ud)$, the nucleon can contain both scalar and vector
diquarks, and the quark $q$ must be properly antisymmetrized with the
identical quark in $(ud)$ to satisfy Fermi
statistics~\cite{Carlson:2005pe}.  Nevertheless,
Fig.~\ref{Fig:Pomeron} illustrates the central point that the Pomeron
carries the entirety of the momentum transfer $q^2$ in the
$t$-channel, suggesting (as is observed) strong forward peaking of the
differential cross section, while a $u$-channel backward peak would
require a complicated intermediate state carrying not only a Pomeron,
but also nonzero baryon number and hidden strangeness.

We propose an additional mechanism for $\phi$ photoproduction, namely,
the production of an $(sq)$ color-$\bar{\bf 3}$ bound diquark $\delta$
and an $[\bar s (ud)]$ color-{\bf 3} bound antitriquark $\bar \theta$,
as in Fig.~\ref{Fig:Exchange}.  In this case, $(ud)$ truly does refer
to a diquark component in $N$, which at first blush can be either of
the ``good'' (scalar-isoscalar) or ``bad'' (vector-isovector) variety,
although data from charge and magnetic nucleon radii prefer the
``good'' component~\cite{Carlson:2005pe}, once proper
antisymmetrization of the wave function between $q$ and $(ud)$ is
performed.  In comparison, in the case of $K\Lambda$ photoproduction
described above, the $\Lambda$ contains only the ``good'' $(ud)$
diquark.  The importance of including diquark baryon substructure in
AdS/QCD models to obtain Regge trajectories matching those of mesons
(as is observed), is emphasized in
Refs.~\cite{deTeramond:2014asa,Dosch:2015nwa}.

Since both $\delta$ and $\bar \theta$ are colored objects,
hadronization of the pair is accomplished as in
Refs.~\cite{Brodsky:2014xia,Lebed:2015tna}, by means of the hadron
wave functions stretching across the space between the colored bound
states.  One can obtain in this way not only a $\phi = (\bar s s)$, $N
= q(ud)$ final state, but also a contribution to the final state $K =
(\bar s q)$, $\Lambda = s(ud)$ expected to be smaller than the one
discussed in the previous section, due to the smaller strength of the
${\bf 3} \otimes {\bf 3} \supset \bar{\bf 3}$ attraction compared to
that of ${\bf 3} \otimes \bar{\bf 3} \supset {\bf 1}$.

Our purpose is not to calculate the complete amplitude for this entire
process ({\it e.g.}, the techniques of Ref.~\cite{Gunion:1973ex} are
useful for the large-$\theta$ region), but only to demonstrate that a
natural physical mechanism exists to provide an interaction producing
an enhanced cross section in both the forward and backward directions.
We can therefore make a number of simplifying assumptions.  First, we
neglect Fermi motion within the $\delta$ and $\bar \theta$, so that
the constituents within each bound state move with the same velocity
(this assumption can be lifted by folding in appropriate distribution
functions; {\it e.g.}, in the light-front formalism, see
Ref.~\cite{Forshaw:2003ki}).  Then it is clear from
Fig.~\ref{Fig:Exchange} that forward $(\theta \to 0)$ scattering of
$(sq)$ also produces a forward-scattered $\phi$, since in the c.m.\
$(sq)$ and $[\bar s (ud)]$ have the same momentum magnitude, but the
former is lighter and therefore has a larger speed.  In turn, $s$ has
a larger speed than $\bar s$.  Since $m_s = m_{\bar s}$, the net
momentum of $\phi = (s\bar s)$ then points in the same direction as
$s$, and hence, as $(sq)$.

This strategy has much in common with the ``hard scattering approach''
(HSA) developed in Refs.~\cite{Efremov:1979qk,Lepage:1980fj}, in that
both take all constituents to move collinear with their parent
hadrons, include the minimal necessary number of gluon exchanges to
accomplish the process, and require folding in appropriate
distribution amplitudes.  The HSA approach applied to $\phi$
photoproduction at high energy scales (with genuinely hard gluons) has
been studied in Ref.~\cite{Goritschnig:2014goa}, although using older
data than here.

\begin{figure}[!ht]
\begin{center}
\includegraphics[width=\linewidth]{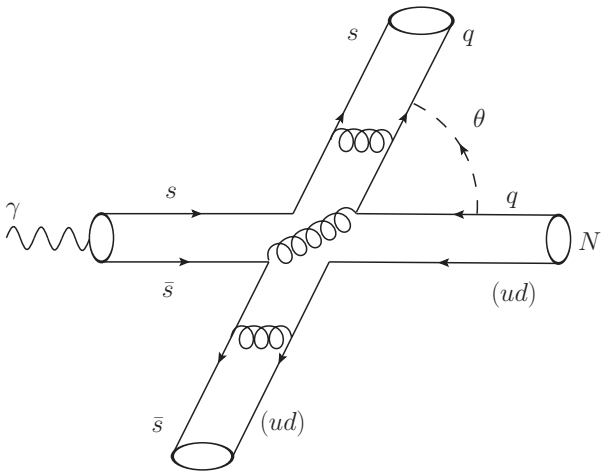}
  \caption{As in Fig.~\ref{Fig:Exchange}, but including a minimal
  gluon exchange that naturally explains the enhancement of $\phi$
  production in the $\theta \to 0$ direction, as described in the
  text.}
\label{Fig:Gluon1}
\end{center}
\end{figure}

Since the diagrams again have four constituents, the minimum number of
gluon exchanges necessary for the process remains three.  Unlike in
Pomeron exchange, however, the $\delta$-$\bar \theta$ production
mechanism provides natural alternatives that produce enhancements in
the forward and backward directions.  Consider first
Fig.~\ref{Fig:Gluon1}; here, the scattering process is driven by the
lightest constituent ($q$) from either of the initial particles ($N$)
exchanging a gluon with the $\bar s$ constituent of the other initial
particle ($\gamma$).
As just discussed, the $\delta$-$\bar \theta$ configuration favored
for forward $\phi$ production is the one in which $(sq)$ is also
produced in the forward direction, recoiling against $[\bar s (ud)]$.
Since the lightest constituent (here, $q$) is the easiest to deflect
through large angles, this particular exchange diagram provides a
natural mechanism for backward scattering of the $q$, as well as that
of the $\bar s$.

Once the $q$ and $\bar s$ are deflected through a large angle, the $s$
and $(ud)$ must each be deflected (through a generically smaller
angle) to become bound to into their respective $(sq)$ and $[\bar s
(ud)]$ combinations.  This binding is accomplished in our crude
picture by the exchange of a gluon between $s$ and $q$, and between
$\bar s$ and $(ud)$, as depicted in Fig.~\ref{Fig:Gluon1}.  As
promised, the minimal number of gluon exchanges required to achieve
the desired final states is precisely three.  One therefore naively
expects this diagram to peak in the forward direction, and in the next
section we see this expectation indeed to be realized.

\begin{figure}[!ht]
\begin{center}
\includegraphics[width=\linewidth]{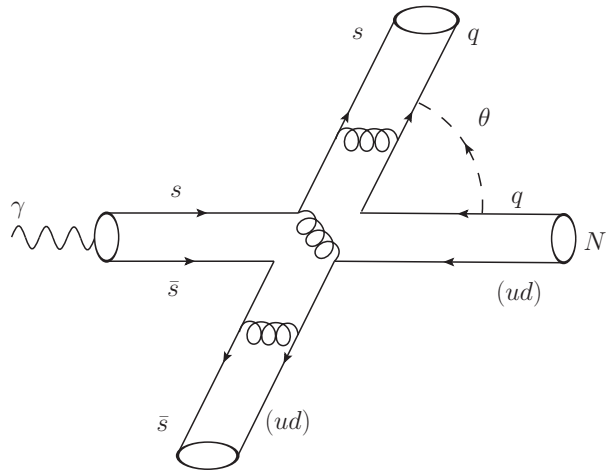}
  \caption{As in Fig.~\ref{Fig:Exchange}, but including a minimal
  gluon exchange that naturally explains the enhancement of $\phi$
  production in the $\theta \to \pi$ direction, as described in the
  text.  The angle $\theta$ exhibited here is identical to the one in
  Fig.~\ref{Fig:Gluon1}, for the purpose of ease of comparison.}
\label{Fig:Gluon2}
\end{center}
\end{figure}

A natural exchange diagram producing a backward peak is depicted in
Fig.~\ref{Fig:Gluon2}.  Here, the interaction gluon connects the
heaviest component [$(ud)$] to the $s$,
and is expected to produce an enhancement when both of them are
deflected in the backward direction, the $\theta \to \pi$ limit of
Fig.~\ref{Fig:Gluon2}.  In that case, in order to form the $[\bar s
(ud)]$ and $(sq)$ combinations, the $\bar s$ and $q$ deflect through a
smaller angle, and binding is accomplished through gluon exchanges
between $(ud)$ and $\bar s$, and between $s$ and $q$, respectively.
Again, a minimum of precisely three gluon exchanges is necessary to
obtain the desired final state.  And since backward deflection of the
heavier $(ud)$ is not as easy as backward deflection of the lighter
$q$, one expects the size of the enhancement in the backward direction
produced by Fig.~\ref{Fig:Gluon2} to be smaller than the enhancement
in the forward direction produced by Fig.~\ref{Fig:Gluon1}, which is
verified in the next section to be true.

Before proceeding to a calculation, we hasten to emphasize its
extremely rudimentary nature.  First, we are investigating a
particular production mechanism ($\delta$-$\bar \theta$) that has
never been demonstrated unambiguously to occur in any process.  The
$(sq)$ and $[\bar s (ud)]$ are assumed to act as bound quasiparticles,
allowing for analysis as a $2 \to 2$ process.  Second, we suppose that
the reaction proceeds through a ``good'' diquark component of the
nucleon, which interacts as a unit and maintains its identity
throughout the process.  Third, we neglect the Fermi motion of each
composite particle and assume that each initial and final state
consists of two collinear components moving at the same
velocity. Fourth, we use the minimum required number of gluons for
each process, despite the fact that the reaction and binding
interactions are most certainly nonperturbative.  No hard scale is
present to justify a perturbative treatment, and the gluons here
should be thought of only as quasiparticles that accomplish momentum
transfer among the components; in reality, these gluons simply
represent collections of multiple gluon exchanges that accomplish the
same overall scattering.  In fact, the specific choices of topology
for the diagrams in Figs.~\ref{Fig:Gluon1}, \ref{Fig:Gluon2}, even
with the minimal gluon exchange number, are not unique.  For example,
while it is tempting to apply the above narrative that the interaction
between some of the constituents occurs first, and then the remaining
constituents exchange gluons with the scattered components to form
bound states, it is just as tenable from a quantum field-theory point
of view that the initial constituents $\gamma$ and $N$ come unbound by
gluon exchange first, and then the interaction occurs.  Or, the
binding and unbinding gluons can stretch across the
interaction\footnote{Such gluons would be nonplanar, and give
contributions formally suppressed by $O(1/N_c^2)$ relative to diagrams
without them, where $N_c = 3$ is the number of QCD colors.}.  In real
QCD, one expects {\em all\/} such gluons to appear---copiously---in a
typical diagram.

\begin{figure}[!ht]
\begin{center}
\includegraphics[width=\linewidth]{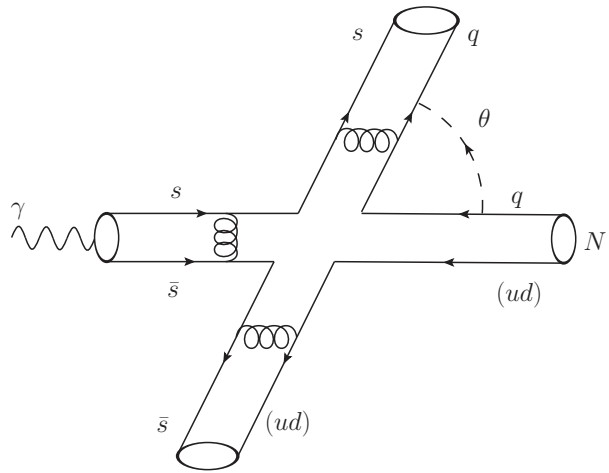}
  \caption{As in Fig.~\ref{Fig:Exchange}, but including a minimal
  gluon exchange that does not have a natural central interaction
  gluon connecting initial and final states.}
\label{Fig:Gluon3}
\end{center}
\end{figure}

To establish a systematic calculation, we consider all 12 diagrams
that have the minimum three planar gluons, such that each of the
constituents must couple at least once to a gluon.
Figures~\ref{Fig:Gluon1} and~\ref{Fig:Gluon2} may be considered as
representatives of diagram classes [with amplitudes ${\cal M}^{(1)}$
and ${\cal M}^{(2)}$, respectively], as defined by the placement of
the central interaction gluon that connects initial and final states.
The members of each class are defined by which two external states
have their constituents connected by a gluon.  In a third allowed
class represented by Fig.~\ref{Fig:Gluon3}
[amplitudes ${\cal M}^{(3)}$], the constituents of each of three of
the external states are connected by a gluon.  The definition of the
12 diagrams is summarized in Table~\ref{Table:Diags}; thus, for
example, the amplitudes corresponding to the literal diagrams of
Figs.~\ref{Fig:Gluon1}, \ref{Fig:Gluon2}, and \ref{Fig:Gluon3} are
${\cal M}^{(1)}_1$, ${\cal M}^{(2)}_1$, and ${\cal M}^{(3)}_1$,
respectively.  It is also simple to check that each of these 12
diagrams also has exactly two internal constituent propagators.

\begin{table}[ht]
  \caption{Amplitudes based on the classes defined by
  Figs.~\ref{Fig:Gluon1}, \ref{Fig:Gluon2}, \ref{Fig:Gluon3} [defined
  as ${\cal M}^{(1),(2),(3)}$, respectively].  Each amplitude is
  defined by its class and the listed subset of initial and final
  states, each member of which exchanges a single binding gluon.}
  \label{Table:Diags}
\begin{tabular}{cc|cc|cc}
\hline
${\cal M}^{(1)}_1$ & ($\delta$, $\bar \theta$) &
${\cal M}^{(2)}_1$ & ($\delta$, $\bar \theta$) &
${\cal M}^{(3)}_1$ & ($\gamma$, $\delta$, $\bar \theta$) \\
${\cal M}^{(1)}_2$ & ($\gamma$, $N$) &
${\cal M}^{(2)}_2$ & ($\gamma$, $N$) &
${\cal M}^{(3)}_2$ & ($N$, $\delta$, $\bar \theta$) \\
${\cal M}^{(1)}_3$ & ($\delta$, $N$) &
${\cal M}^{(2)}_3$ & ($\delta$, $\gamma$) &
${\cal M}^{(3)}_3$ & ($\gamma$, $N$, $\delta$) \\
${\cal M}^{(1)}_4$ & ($\gamma$, $\bar \theta$) &
${\cal M}^{(2)}_4$ & ($N$, $\bar \theta$) &
${\cal M}^{(3)}_4$ & ($\gamma$, $N$, $\bar \theta$) \\
\hline
\end{tabular}
\end{table}

\section{Calculation and Results} \label{sec:Results}

We estimate the relative size of the diagrams in
Table~\ref{Table:Diags} by establishing kinematics and definitions of
momenta, and then by calculating the angular dependence originating
from the product of momentum transfer factors forming the denominators
of the three gluon propagators and two constituent propagators
appearing in each diagram.  Spin structures in the form of Dirac
matrices or Lorentz tensors are also ignored; indeed, the issue of how
to parametrize the coupling of a gluon to the $(ud)$ diquark is
avoided with this assumption.  The precise recipe for treating masses
appearing in these factors is described below.  In short, we estimate
the relative sizes and angular dependences of the 12 amplitudes ${\cal
  M}$ in Table~\ref{Table:Diags} by using just a few of the factors
appearing in each of them.

We begin with kinematical conventions.  The external composite
particle masses, not only $m_\gamma = 0$ and $m_p$, but also the
diquark $m_{(sq)}$ and antitriquark $m_{[\bar s(ud)]}$ masses, are
assumed known, uniquely determining the c.m.\ energy and momenta of
the external particles in terms of the total c.m.\ energy $\sqrt{s}$
of the process, as in any $2 \to 2$ process:
\begin{eqnarray}
\sqrt{s} = E_{\rm tot}^{\rm c.m.} & = & \sqrt{m_N^{\vphantom{2}}
(2E_{\gamma}^{\rm lab} + m_N^{\vphantom{2}})} \, , \nonumber \\
E_{\gamma}^{\rm c.m.} & = & \frac{s - m_N^2}{2\sqrt{s}} \, , \nonumber
\\
E_N^{\rm c.m.} & = & \frac{s + m_N^2}{2\sqrt{s}} \, , \nonumber \\
p_{\gamma}^{\rm c.m.} = p_N^{\rm c.m.} & = & E_{\gamma}^{\rm c.m.} \,
, \nonumber \\
E_{(sq)}^{\rm c.m.} & = & \frac{s - m_{[\bar s (ud)]}^2 + m_{(sq)}^2}
{2\sqrt{s}}
\, , \nonumber \\
E_{[\bar s (ud)]}^{\rm c.m.} & = & \frac{s - m_{(sq)}^2 +
m_{[\bar s (ud)]}^2}{2\sqrt{s}} \, ,
\end{eqnarray}
and
\begin{eqnarray}
\lefteqn{p_{(sq)}^{\rm c.m.} = p_{[\bar s (ud)]}^{\rm c.m.} =}
\nonumber \\
& & {\frac{\sqrt{ \left[ s - \left( m_{[\bar s (ud)]} \! + m_{(sq)}
\right)^2 \right] \left[ s - \left( m_{[\bar s (ud)]} \! - m_{(sq)}
\right)^2 \right]}} {2\sqrt{s}} \, .} \nonumber \\
\end{eqnarray}

Since the constituents of each composite particle are assumed to move
with the same velocity, the fraction of the total momentum carried by
each is assumed to be the ratio of its mass to the total mass of the
constituents.  In particular, if we define
\begin{eqnarray}
r_1 & \equiv & \frac{m_s}{m_s + m_{\bar s}} \, , \nonumber \\
r_2 & \equiv & \frac{m_q}{m_q + m_{(ud)}} \, , \nonumber \\
r_3 & \equiv & \frac{m_q}{m_s + m_q} \, , \nonumber \\
r_4 & \equiv & \frac{m_{\bar s}}{m_{\bar s} + m_{(ud)}} \, ,
\label{eq:mass_ratios}
\end{eqnarray}
and denote initial and final constituent momenta with primes on the
latter and not on the former, then
\begin{eqnarray}
p_s               & = & r_1 \, p_\gamma \, , \nonumber \\
p_{\bar s}        & = & (1-r_1) p_\gamma \, , \nonumber \\
p_q               & = & r_2 \, p^{\vphantom{2}}_N \, , \nonumber \\
p_{(ud)}          & = & (1-r_2) p^{\vphantom{2}}_N \, , \nonumber \\
p^\prime_s        & = & (1-r_3) p_{(sq)} \, , \nonumber \\
p^\prime_{\bar s} & = & r_4 \, p_{[\bar s (ud)]} \, , \nonumber \\
p^\prime_q        & = & r_3 \, p_{(sq)} \, , \nonumber \\
p^\prime_{(ud)}   & = & (1-r_4) p_{[\bar s (ud)]} \, .
\label{eq:mom_defns}
\end{eqnarray}
This simple apportionment of momenta among constituents is completely
analogous to the result in light-front quantum field
theory~\cite{Brodsky:1997de}.  Conservation of 4-momentum at each
vertex then completely determines the momentum transfers in terms of
the $r_i$ factors and scalar products between the external momenta,
which in turn are completely determined by the external particle
masses and the scattering angle $\theta$.

Out of the 12 diagrams, only six distinct gluon momenta appear:
\begin{eqnarray}
q_s         & \equiv & p^\prime_s - p_s \, , \nonumber \\
q_{\bar s}  & \equiv & p^\prime_{\bar s} - p_{\bar s} \, \nonumber \\
q_q         & \equiv & p^\prime_q - p_q \, , \nonumber \\
q_{(ud)}    & \equiv & p^\prime_{(ud)} - p_{(ud)} \, , \nonumber \\
q_{(sq)}    & \equiv & p_{(sq)} - p_s - p_q = p_{\bar s} + p_{(ud)} -
p_{[\bar s (ud)]} \, , \nonumber \\
q_{s\bar s} & \equiv & p^\prime_s + p^\prime_{\bar s} - p_\gamma =
p^{\vphantom{2}}_N - p^\prime_q - p^\prime_{(ud)} \, ,
\end{eqnarray}
which satisfy between them three simple identities:
\begin{eqnarray}
q_{(sq)} & = & q_s + q_q = - q_{\bar s} - q_{(ud)} \, , \nonumber \\
q_{s\bar s} & = & q_s + q_{\bar s} = - q_q - q_{(ud)} \, , \nonumber
\\
0 & = & q_s + q_{\bar s} + q_q + q_{(ud)} \, ,
\end{eqnarray}
so that only three of them are linearly independent, exactly as one
expects for momentum transfers between four independent external
momenta.  Compared to the full Feynman calculation with gluons in the
amplitudes ${\cal M}$, we treat the gluon propagator factors $q_j^2$
as the only relevant ones for this analysis ({\it i.e.}, we neglect
polarization tensor structures).  However, using skeletal diagrams
such as Figs.~\ref{Fig:Gluon1}, \ref{Fig:Gluon2}, \ref{Fig:Gluon3} to
model much more complicated diagrams with extensive gluon and internal
$q\bar q$ exchanges introduces the potential for artificial kinematic
singularities to arise when some of the internal lines go on mass
shell.  Part of this problem is caused by the assumption of zero
transverse constituent momenta, but much of it simply is due to the
usual soft-gluon infrared singularities one expects in a perturbative
treatment.  Physically, an on-shell gluon corresponds to an
infinite-range interaction between the constituents, which does not
comport with the finite range associated with confinement.  To account
for this important dynamics, we provide the gluons with a finite mass
scale $m_{\rm conf}$ (with a specific value chosen below), and
introduce dimensionless gluon propagator factors
\begin{equation} \label{eq:gluon_prop}
Q_j \equiv -\frac{m_{\rm conf}^2}{q_j^2 - m_{\rm conf}^2} \, ,
\end{equation}
where the sign makes $Q_j$ positive in the usual case of scattering
four-momenta.

Furthermore, eight distinct internal constituent momenta (in
which the subscript refers to the particular constituent line) appear:
\begin{eqnarray}
k_s               & \equiv & p_{(sq)} - p_q \, , \nonumber \\
k_{\bar s}        & \equiv & p_{[\bar s (ud)]} - p_{(ud)} \, ,
\nonumber \\
k_q               & \equiv & p_{(sq)} - p_s \, , \nonumber \\
k_{(ud)}          & \equiv & p_{[\bar s (ud)]} - p_{\bar s} \, ,
\nonumber \\
k^\prime_s        & \equiv & p_\gamma - p^\prime_{\bar s} \, ,
\nonumber \\
k^\prime_{\bar s} & \equiv & p_\gamma - p^\prime_s \, , \nonumber \\
k^\prime_q        & \equiv & p^{\vphantom{2}}_N - p^\prime_{(ud)} \, ,
\nonumber \\
k^\prime_{(ud)}   & \equiv & p^{\vphantom{2}}_N - p^\prime_q \, ,
\label{eq:constit_mom}
\end{eqnarray}
which are linearly independent except for the overall momentum
conservation constraint.  In the full amplitudes ${\cal M}$, the
factors $\bar u_j (p_f) \gamma^\mu (\slash \! \! \! k_j - m_j)^{-1}
\gamma^\nu u_j (p_i)$ appear for the fermionic constituents.  By treating
the bosonized propagator denominator $(k_j^2 - m_j^2)$ as the only
significant factor for this analysis, {\it i.e.}, by ignoring the
effect of the spinor algebra except to use the Dirac equation of
motion to eliminate $\slash \! \! \! k_j$ from the numerator, by using
the conventional $2m_j$ normalization for spinors, and by noting that
scattering momentum transfers are negative, the relevant dimensionless
factors for constituent lines in this analysis are:
\begin{equation} \label{eq:constit_prop}
K^{(\prime)}_j \equiv -\frac{2m_j^2}{k^{(\prime) \, 2}_j - m_j^2} \, .
\end{equation}
For convenience, we employ the same form for the (bosonic) diquark
constituent $(ud)$.  From standard relativistic kinematics using
Eqs.~(\ref{eq:mass_ratios})--(\ref{eq:constit_prop}), one finds that
the propagator factors $Q_s$, $Q_{(ud)}$, $Q_{(sq)}$, $Q_{s \bar s}$,
$K_{\bar s}$, $K_q$, $K^\prime_{\bar s}$, and $K^\prime_q$ are
enhanced in the forward direction, while $Q_{\bar s}$, $Q_q$, $K_s$,
$K_{(ud)}$, $K^\prime_s$, and $K^\prime_{(ud)}$ are enhanced in the
backward direction.  These results for $Q_s$, $Q_{(ud)}$, $Q_{\bar
  s}$, $Q_q$ were anticipated in the last section.

The choice of appropriate constituent masses also requires some care.
Current quark masses would only be appropriate in a fully perturbative
analysis in which classes of diagrams are resummed to avoid
singularities associated with particles going on shell.  However,
traditional constituent masses are not entirely appropriate for this
calculation either, as they are typically obtained from static
processes, not a dynamical scattering such as photoproduction.  The
choice of mass parameters, as seen below, lies somewhere in between
these extremes.

Even then, the appropriate mass to choose for a given internal
constituent line suffers from ambiguity.  Take, for example, the
factor $k_q$ in Eq.~(\ref{eq:constit_mom}), which refers to the
momentum of a light $q$ emerging from the breakup of the nucleon $N$
{\it en route\/} to binding into a diquark $(sq)$.  According to
Eq.~(\ref{eq:mom_defns}), should its mass be considered a fraction
$r_2$ of $m_N$, or a fraction $r_3$ of $m_{(sq)}$?  Or indeed, since
the propagator lies deep in the diagram, should one use the
confinement mass scale $m_{\rm conf}$ previously introduced?  We adopt
the convention that the appropriate effective constituent mass to
appear in the propagators Eq.~(\ref{eq:constit_prop}) is the maximum
of the two suggested by the initial ($i$) and final ($f$) states into
which it binds [according to Eqs.~(\ref{eq:mass_ratios}) and
(\ref{eq:mom_defns})] {\em plus\/} a contribution from the confinement
scale $m_{\rm conf}$, to account for the expected off-shell behavior:
\begin{equation}
m_{j , \rm eff}^2 = \max ( m_{j,i}^2 , m_{j,f}^2 ) + m_{\rm conf}^2 \,
.
\end{equation}

The model amplitudes of Table~\ref{Table:Diags} in the simplified
notation of Eqs.~(\ref{eq:gluon_prop}) and (\ref{eq:constit_prop})
read
\begin{eqnarray}
{\cal M}^{(1)}_1 & = & Q_s Q_{(ud)} Q_{(sq)} K_{\bar s} K_q \, ,
\nonumber  \\
{\cal M}^{(1)}_2 & = & Q_s Q_{(ud)} Q_{s \bar s} K^\prime_{\bar s}
K^\prime_q \, , \nonumber \\
{\cal M}^{(1)}_3 & = & Q_s Q_{(ud)} Q_{\bar s} K_q K^\prime_q \, ,
\nonumber \\
{\cal M}^{(1)}_4 & = & Q_s Q_{(ud)} Q_q K_{\bar s} K^\prime_{\bar s}
\, , \nonumber \\
{\cal M}^{(2)}_1 & = & Q_{\bar s} Q_q Q_{(sq)} K_s K_{(ud)} \, ,
\nonumber \\
{\cal M}^{(2)}_2 & = & Q_{\bar s} Q_q Q_{s \bar s} K^\prime_s
K^\prime_{(ud)} \, , \nonumber \\
{\cal M}^{(2)}_3 & = & Q_{\bar s} Q_q Q_{(ud)} K_s K^\prime_s \, ,
\nonumber \\
{\cal M}^{(2)}_4 & = & Q_{\bar s} Q_q Q_s K_{(ud)} K^\prime_{(ud)} \,
, \nonumber \\
{\cal M}^{(3)}_1 & = & Q_q Q_{(ud)} Q_{(sq)} K_s K_{\bar s} \, ,
\nonumber \\
{\cal M}^{(3)}_2 & = & Q_s Q_{\bar s} Q_{(sq)} K_q K_{(ud)} \, ,
\nonumber \\
{\cal M}^{(3)}_3 & = & Q_{\bar s} Q_{(ud)} Q_{s \bar s} K^\prime_s
K^\prime_q \, , \nonumber \\
{\cal M}^{(3)}_4 & = & Q_s Q_q Q_{s \bar s} K^\prime_{\bar s}
K^\prime_{(ud)} \, . \label{eq:Amps}
\end{eqnarray}
The chosen input masses, all in MeV, are listed in
Table~\ref{Table:Inputs}.  Again, guidance for this choice is
suggested, but not determined, by an interpolation between typical
current and constituent masses used in the literature.  For example,
one can make an argument for $(ud)$ diquark masses anywhere from
$\simeq 30$~MeV ({\it i.e.}, not many times more than the sum of
current quark masses) up to $\simeq 600$~MeV ({\it i.e.}, two-thirds
of a nucleon, or the mass of a $\sigma$ meson).
\begin{table}[ht]
  \caption{Input masses in MeV.}
  \label{Table:Inputs}
\begin{tabular}{c|c|c|c|c|c|c|c|c}
$\sqrt{s}$ & $m_\gamma$ & $m_N$ & $m_{(sq)}$ & $m_{[\bar s (ud)]}$ &
$m_{(ud)}$ & $m_s = m_{\bar s}$ & $m_q$ & $m_{\rm conf}$ \\
\hline
2200 & 0 & 938.3 & 680 & 1150 & 100 & 200 & 10 & 425 \\
\end{tabular}
\end{table}
The most important inputs in obtaining a result resembling
experimental data appear to be choosing $m_{(sq)} + m_{[\bar s (ud)]}$
not terribly far below the observed forward peak at $\sqrt{s} =
2.2$~GeV, and choosing a fairly large value for $m_{\rm conf}$.  The
necessary magnitude of $m_{\rm conf}$ appears to arise largely due to
the feature of this simplified model that the initial $s$ and $\bar s$
quarks from the dissociation of $\gamma$ are lightlike, as seen from
the first two of Eq.~(\ref{eq:mom_defns}); presumably, introducing
substantial transverse momenta in a more realistic calculation
produces the same effect as $m_{\rm conf}$.  In any case, the model
presented here is so simple that the input masses used should be
viewed only as a qualitative guide to obtaining results in accord with
experiment.

\begin{figure}[!ht]
\begin{center}
\includegraphics[width=\linewidth]{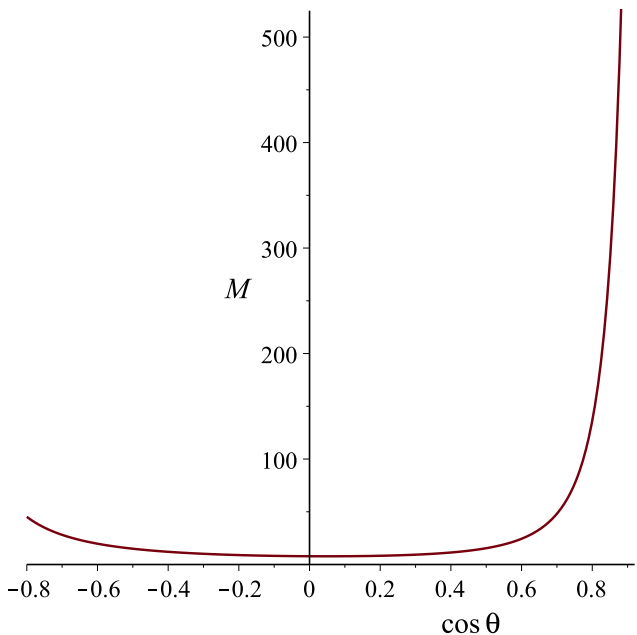}
  \caption{Coherent sum $M$ of the 12 amplitudes of
  Eq.~(\ref{eq:Amps}) using the inputs in Table~\ref{Table:Inputs}, as
  a function of $\cos \theta$.}
\label{Fig:Amp_Sum}
\end{center}
\end{figure}

For the given inputs, the amplitudes of class ${\cal M}^{(1)}$ are all
strongly peaked in the forward direction, while those of class ${\cal
M}^{(2)}$ are all strongly peaked in the backward direction but
possess in addition a smaller forward peak.  In particular, the
amplitude ${\cal M}^{(1)}_1$ [${\cal M}^{(2)}_1$] corresponding to
Fig.~\ref{Fig:Gluon1} (Fig.~\ref{Fig:Gluon2}) shows preferential
forward (backward) scattering, as was anticipated from general
arguments.  The diagrams of class ${\cal M}^{(3)}$ also turn out to be
forward enhanced, but many times smaller than those in class ${\cal
M}^{(1)}$.  To illustrate the forward-backward peaking, we present in
Fig.~\ref{Fig:Amp_Sum} the simple coherent sum $M$ of the 12
amplitudes.  Obviously, the absence of numerous factors in the full
Feynman amplitudes---not least of which are relative signs leading to
destructive interferences---means that the plot of
Fig.~\ref{Fig:Amp_Sum} should only be taken seriously in its coarsest
features, namely, a large peak in the forward direction and a small
peak in the backward direction.  The particular value for the
forward-to-backward ratio is simple to adjust using slightly different
inputs.  Furthermore, we note that the CLAS data presented in
Ref.~\cite{Dey:2014tfa} does not extend to the most forward and
backward directions; the limitation $-0.80 \le \cos \theta \le +0.92$
is used in Fig.~\ref{Fig:Amp_Sum}.  The literal ratio $M(\cos \theta =
+0.92)/M(\cos \theta = -0.8)$ in Fig.~\ref{Fig:Amp_Sum} is about 30\@.

\begin{figure}[!ht]
\begin{center}
\includegraphics[width=\linewidth]{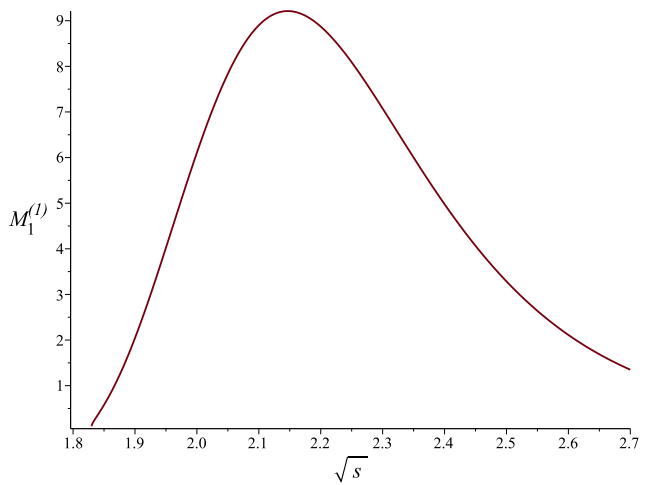}
\caption{Behavior of the amplitude ${\cal M}^{(1)}_1$ as a function of
  $\sqrt{s}$ (in GeV), with the other inputs as in
  Table~\ref{Table:Inputs}.}
\label{Fig:s_behavior}
\end{center}
\end{figure}

Lastly, one may ask whether the forward enhancement behaves away from
the particular chosen peak value $\sqrt{s} = 2.2$~GeV similarly to its
appearance in the data.  To illustrate the result, we plot in
Fig.~\ref{Fig:s_behavior} the amplitude $M^{(1)}_1$ as a function of
$\sqrt{s}$ but otherwise use the mass inputs of
Table~\ref{Table:Inputs}.  The amplitude, rather than its square, is
relevant because its chief contribution to data would presumably
appear through interference with the dominant Pomeron amplitude.  The
full width at half maximum appears to extend from just below 2.0~GeV
to about 2.4~GeV, which is very much like the experimental result
presented in Refs.~\cite{Dey:2014tfa,Dey:2014npa}.  In this model, the
enhancement is indeed due to a special correlated five-quark
configuration, but it is not a literal resonant pentaquark state.
That is, the anomalies may be said to correspond to ``would-be''
pentaquarks.

\section{Conclusions} \label{sec:Concl}

We have seen that interesting forward-backward enhancements observed
in recent data for the photoproduction process $\gamma p \to \phi p$
can easily be explained through the production of a $(su)$
color-$\bar{\bf 3}$ bound diquark and an $[\bar s (ud)]$ color-{\bf 3}
bound antitriquark, using the preferential color couplings ${\bf 3}
\otimes {\bf 3} \supset \bar {\bf 3}$.  Such a configuration (with $s
\to c$) has previously been advocated as an explanation of the exotic
charmoniumlike states.  Here, however, no claim is made that the
enhancements are the result of true resonances, but they are predicted
to create $s$-dependent bumps in the data.

The model relies only on the preference of a system with largely
collinear constituents to minimize the constituent momentum transfers.
Essentially no dynamics is included; the minimal set of gluon
exchanges used for each diagram does not incorporate color physics in
any fundamental way.  Likewise, the constituents are assumed at any
point in the diagrams to carry only momenta parallel to the bound
state in which they occur.  Nevertheless, using plausible values for
constituent masses, one obtains results in good qualitative, and even
semi-quantitative, agreement with experiment.  Numerous possible
improvements leading to stable and predictive results for this process
should certainly be undertaken.

The possibility that diquark/triquark structure can be discerned in
lighter quark systems has long been discussed, but the experimental
signals have always been tantalizingly vague.  In the analogues to the
types of experiments already performed or proposed for heavy-quark
exotics, one can hope to find a clear indication for these novel
structures.

\begin{acknowledgments}
  I gladly acknowledge M.~Karliner for a conversation on his recent
  paper, which ultimately inspired the idea for this work.  I also
  thank S.~Brodsky for providing perceptive comments, M.~Dugger for
  generating detailed plots of CLAS data, and the CERN Theoretical
  Physics Group for support during this work's completion.  This work
  was supported by the National Science Foundation under Grant Nos.\
  PHY-1068286 and PHY-1403891.
\end{acknowledgments}


\end{document}